\documentclass[runningheads]{llncs}

\usepackage[T1]{fontenc}
\usepackage{graphicx}
\usepackage{hyperref}
\usepackage{xcolor}
\hypersetup{
    colorlinks,
    linkcolor={red!50!black},
    citecolor={blue!50!black},
    urlcolor={blue!80!black}
}
\usepackage[nameinlink]{cleveref}
\usepackage{longtable}
\usepackage[frozencache,cachedir=.]{minted}
\setminted{frame=lines,
    framesep=2mm,
    fontsize=\footnotesize,
    breaklines=true,
    linenos
    }

\begin{document}

\title{Implementing ISO/IEC TS 27560:2023 Consent Records and Receipts for GDPR and DGA}

\author{Harshvardhan J. Pandit\inst{1,2}\orcidID{0000-0002-5068-3714
} \and
Jan Lindquist\inst{3} \and
Georg P. Krog\inst{4}}

\titlerunning{Implementing ISO-27560 Consent Records and Receipts for GDPR and DGA}
\authorrunning{H. J. Pandit, J. Lindquist, and G. P. Krog}

\institute{ADAPT Centre, Dublin City University, Dublin, Ireland \email{me@harshp.com} \and
Cybersecurity and Data Protection Group, National Standards Institute, Ireland \and
Privacy and Security Group, Institute for Standards, Sweden \email{jan@linaltec.com} \and 
Signatu AS, Oslo, Norway \email{georg@signatu.com}}
\maketitle              
\begin{abstract}
The ISO/IEC TS 27560:2023 Privacy technologies — Consent record information structure provides guidance for the creation and maintenance of records regarding consent as machine-readable information. It also provides guidance on the use of this information to exchange such records between entities in the form of 'receipts'. In this article, we compare requirements regarding consent between ISO/IEC TS 27560:2023, ISO/IEC 29184:2020 Privacy Notices, and the EU's General Data Protection Regulation (GDPR) to show how these standards can be used to support GDPR compliance. We then use the Data Privacy Vocabulary (DPV) to implement ISO/IEC TS 27560:2023 and create interoperable consent records and receipts. We also discuss how this work benefits the the implementation of EU Data Governance Act (DGA), specifically for machine-readable consent forms.

\end{abstract}

\section{Introduction}\label{sec:intro}
\emph{Informed Consent} is an important legal basis as it provides
control and empowerment to data subjects or users based on the ability
to choose and make decisions. Privacy and data protection laws such as EU's GDPR
\cite{Regulation_GDPR} regulate this process by defining conditions for when
consent should be considered \emph{Valid Consent}. The process of
\emph{Informed Consent} requires information be provided in the form of
a \emph{Consent Notice} to inform the data subject about the processing
that will occur based on the consent and to enable them to make an
\emph{informed choice or decision}.

In order to assess whether an instance of given consent is valid thus
requires keeping records of information regarding how the consent was
obtained i.e. using the notice, and how the consent is being utilised
i.e. the processing enabled through that consent. This same information
is also required for the organisation to determine whether its
processing activities should continue, e.g. depending on whether a
particular user has given consent and whether it is still valid i.e.
hasn\textquotesingle t expired or wasn\textquotesingle t withdrawn).
Such information that is documented and maintained regarding consent is
called a \emph{Consent Record}.

ISO/IEC TS 27560:2023 Consent record information structure \cite{ISO27560} is a Technical
Specification that "specifies an interoperable, open and
extensible information structure" for recording the data
subject\textquotesingle s consent to processing of their personal data
i.e. as consent records, and to provide this information i.e. as consent
receipts. The specification lists \emph{information fields} that
represent specific information associated with consent, and requirements
over the form this information can take e.g. format, number of values,
and whether it is mandatory or optional. It complements the earlier ISO/IEC 29184:2020 Online privacy notices and consent \cite{ISO29184} which describes the information to be provided within privacy notices.

A ISO-27560 conformant implementation fulfils requirements by either storing information directly in the form prescribed by ISO-27560 or by storing information in a form
that can be used to obtain this information. ISO-27560 allows flexibility in how the fields are represented to suit and match domain-specific labels or descriptions, or to introduce additional fields or information types that are needed. 
Such changes, expressed as \emph{schemas} or \emph{profiles}, are still
required to be compatible with the requirements of ISO-27560 such as by requiring the same fields to be mandatory. In this manner, ISO-27560 provides a common, interoperable, and extensible structure for the exchange of information associated with consent. 

In this article, we present an analysis of the requirements for recording consent within ISO-27560 and ISO-29184 and compare them with the requirements for valid consent under GDPR (\cref{sec:comparison}). We then present our work in implementing ISO-27560 using the Data Privacy Vocabulary (DPV) \cite{panditCreatingVocabularyData2019,panditDataPrivacyVocabulary2024} to create a machine-readable, interoperable, and extensible specification for consent records and receipts based on open standards (\cref{sec:dpv}). Through this work we demonstrate the applicability and usefulness of ISO-27560 in assisting with the obligation for demonstrating consent under GDPR (Art.7-1), and explore how ISO-27560 and ISO-29184 can work within the legal framework of GDPR and DGA and the possibility for using this standard to inform the implementation of machine-readable common consent forms under the DGA (\cref{sec:gdpr-dga}). We also discuss practicalities for implementations regarding trust and security (\cref{sec:considerations-security}) and using records and receipts with eIDAS and EUDI wallets (\cref{sec:considerations-wallets}).

\section{Overview of ISO/IEC TS 27560:2023}\label{sec:overview}
\subsubsection{Goals \& Scope}\label{goals-scope}

ISO-27560 has two broad goals: to guide the recording
of information about consent for processing of personal data in a form
that is interoperable, open, and extendable, and to provide information
to individuals. To implement this, it defines several requirements (as
\emph{controls} in ISO terminology) for ensuring the required
information is maintained and is supported by appropriate processes
within the organisation. ISO-27560 is stated as a supplement
to the earlier ISO-29184, where ISO-29184
defines how information is provided via notices in order to request
consent, and ISO-27560 defines how information is recorded
for given consent and provided back to the individual (as receipts).

The objective of ISO-27560 is to define an information
structure for consent record which contains: (1) Information about the
processing of personal data; (2) Privacy notices where this information
was provided; (3) How data was obtained; and (4) Events related to consent
(giving, withdrawing, etc.). It also defines an information structure
providing all or some of this information to the data subject in the
form of a consent receipt. To support implementations, Annex A provides
examples of consent records and receipts using DPV, and
Annex B provides an overview of the different states or stages in
\textquotesingle consent lifecycle\textquotesingle{} - which is based on DPV\textquotesingle s consent states \cite{panditCreatingVocabularyData2019,panditDataPrivacyVocabulary2024} and analysis of existing approaches \cite{kurtevaConsentLensSemantics2021,esteves2022analysis}.

Specific guidance on implementation such as the choice of technologies
is not in the scope of ISO-27560, though its Annexes provide
informative guidance on related topics. Annex C describes performance
and efficiency considerations, Annex D describes format and encoding
structures, Annex E describes security of records and receipts, and
Annex G describes application in Privacy Information Management Systems
(PIMS). Further uses of consent records or receipts, such as how data
subjects can obtain consent receipts or maintain their own consent
records is not described in ISO-27560.

\subsubsection{Consent Records}\label{consent-records}

ISO-27560 defines \emph{Consent Record} as the
documentation of information about a data subject\textquotesingle s
consent for the processing of their personal data in terms of the
details about the processing as well as the interactions related to
consent (e.g. giving or withdrawing it). Consent Records are an
essential part of keeping records regarding whether consent has been
obtained and is valid for processing, and to keep this information for
correctly conducting processing relying on it. ISO-27560 as
well as regulatory requirements such as GDPR Article 7-1
require maintaining consent records where consent is used as the legal
basis. While GDPR Article 7-1 only states that consent should be
demonstrable, ISO-27560 provides an information structure
for how this information should be maintained and what processes should
exist within an organisation in for this.

It is important to distinguish between a \emph{Consent Record} with
several relevant but distinct concepts. A consent record only refers to
the information recorded regarding consent, whereas a \emph{Consent
Notice} refers to the notice and information provided to the data
subject in order to inform them about the processing - such as while
requesting consent. While there is a significant overlap between a
consent record and a consent notice, there are key differences such as
notices orienting information for human consumption (e.g. layering of
dialogues to provide summaries and detailed descriptions) and dictating
the manner in which consent is expressed (e.g. checkboxes for options
and confirmation by clicking a button). In contrast, a consent record is
not required to accurately reflect the manner in which this information
was presented to the user, but to only record it in a manner that
enables assessing whether the consent is given and if so for which
processing activities.

This distinction is evident in ISO-29184 being the
standard for consent notices - which specifies what information should
be present in a notice and the manner in which it is presented. In turn,
ISO-27560 only refers to notices to limit its scope to
representing information necessary within a consent record. Therefore a
consent record, despite containing a link to the notice, is not by
itself sufficient to determine the \emph{validity} of consent, and
instead acts as the primary record containing information or links to
information for conducting such assessments. Its primary purpose is
therefore limited to supporting claims for who is the subject, who is the controller, what is the consent about (e.g. which purpose, what recipients), 
what is the state of consent (e.g. request, given, terminated), and where/when/how it occurred (e.g. accepted on specific timestamp).

A ISO-27560 conformant consent record typically has the
following sections representing relevant information:

\begin{enumerate}
\item
  metadata about the consent record such as its identifier
\item
  the individual associated with the consent i.e. data subject
\item
  the subject of consent i.e. specifics of the processing of personal
  data such as purposes, services, data categories, and storage conditions
\item
  entities involved e.g. data controller and third parties
\item
  relevant contextual information e.g. notice, rights, restrictions
\item
  provenance of events associated with the consent e.g. given, withdrawn
\end{enumerate}

Under GDPR, the obligation to maintain records of consent is explicitly
stated in Article 7-1 and Recital 42. This information includes, at a
minimum, the identity of the Controller and the purposes of processing
(Recital 42). Further, Articles 13 and 14 dictate the information that
must be provided to the data subject which includes recipients,
transfers to third countries, data storage periods and conditions,
existence of rights (including consent withdrawal), and specific
information regarding processing such as the use of automated decision
making or profiling.

\subsubsection{Consent Receipts}\label{consent-receipts}

ISO-27560 defines a consent receipt as an authoritative
document that is used to communicate the existence of a consent record
or to provide information contained within it. It is effectively an
\textquotesingle authoritative copy\textquotesingle{} of a consent
record provided by one entity to another, where it may contain all or only some
information from the consent record regarding the consent
and its relevance to processing activities. Such receipts are useful
to communicate the existence of consent decisions, and enable entities to
exercise of rights or raise issues and complaint regarding processing
activities.

Consent receipts are a relatively newer and under-utilised practice,
with no legal requirements existing that refer to the concept (of
receipts) or state how they should be used. In addition, the usefulness of receipts as information provided to another entity requires consideration of specific terms and norms particular to the domain or sector. ISO-27560
follows this by providing the flexibility for organisations to choose a suitable schema for their particular domain or use-case. It defines a minimal structure consisting of some fields
representing the receipt metadata, but does not have any requirements on
the information structuring within the receipt or its correspondence to fields
within the record.

A ISO-27560 conformant consent receipt only requires a
metadata section providing information about the consent receipt such as
its identifier. Deciding on which additional information is to be
provided and in what forms and using which structures is left up to
implementing entities. In this guide, we presume that the consent
receipt is intended for providing a copy of all information within the
consent record.

According to ISO-27560, records are generated and maintained
by organisations (Controller, Third Party), and are utilised to provide
receipts to a Data Subject. In contrast, the Kantara Consent Receipts
specification {[}ref{]}, upon which ISO-27560 is based,
defines Consent Receipts as being provided by a Data Subject to a
Controller.

For practical considerations of this work, we make no presumptions or
enact restrictions on the use of records and receipts. Any entity, be it
a Controller or a Data Subject, can maintain their own consent record or
issue receipts. Though the phrasing of some sections may imply the
Controller as the implementing entity, it does not preclude another
entity from also implementing ISO-27560.

\subsubsection{Structure}\label{structure}

A Consent Record contains four sections as described below and depicted
visually in Figure \ref{fig:27560} (the terms used are based on the implementation of ISO-27560 for GDPR using DPV as described later in the article):
\begin{figure}
    \centering \label{fig:27560}
    \fbox{\includegraphics[width=\textwidth]{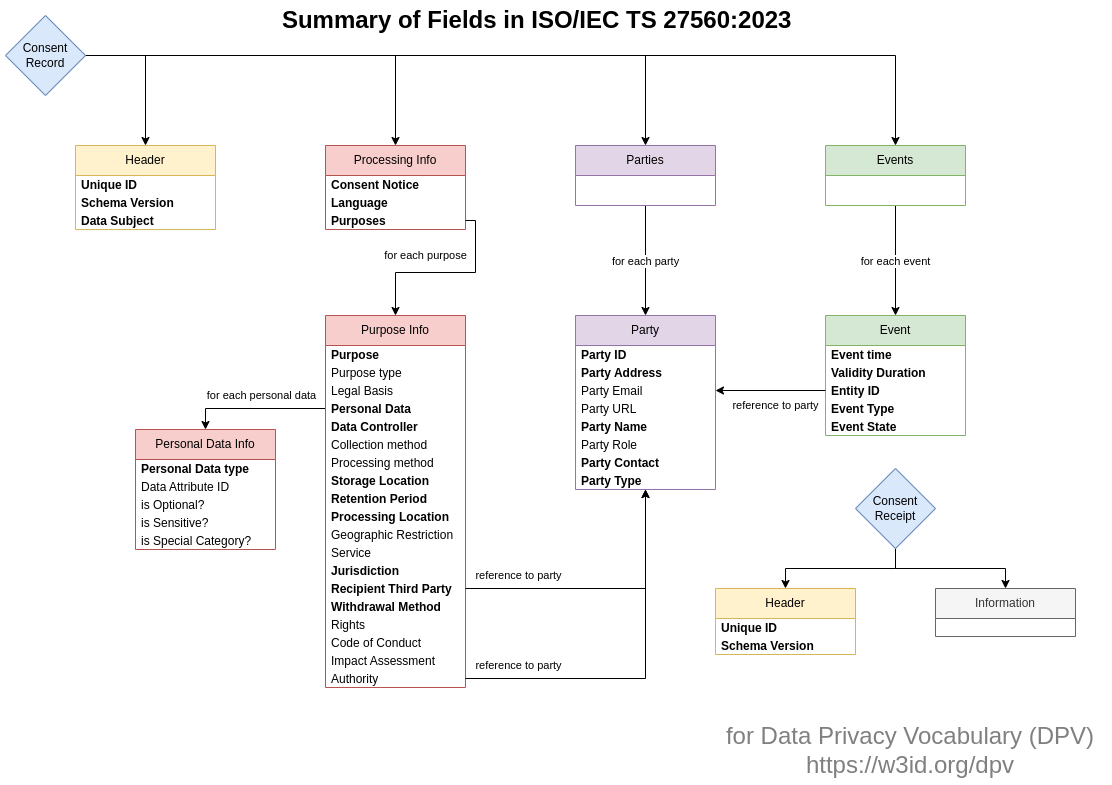}}
    \caption{Summary of fields in ISO/IEC TS 27560:2023. The field names
    have been modified for alignment with DPV concepts. Field names in bold
    are mandatory.}
\end{figure}

\begin{enumerate}
\item
  \textbf{Header Fields:} these provide metadata about the record e.g. its unique identifier and timestamp of creation. These fields also include information on the schema which dictates how the information in the record is structured and which fields are necessary/optional. ISO-27560 permits creation of different schemas to support varying use-cases and information requirements.
\item
  \textbf{Processing Fields:} these provide information about the processing activities e.g. purposes, personal data, storage durations, geographic locations and restrictions, link to privacy notice, rights, and others.
\item
  \textbf{Parties Fields:} these provide information about entities involved in the
  processing e.g. controllers, third parties, authorities. The party has an identifier which is used to link or associate it with fields in the processing section e.g. to indicate which party is the controller.
\item
  \textbf{Events Fields:} these provide information about \emph{consent events} e.g. consent given, consent withdrawn. Information includes the type of event, time, duration, associated entity, and how it was expressed.
\end{enumerate}

Each section contains fields which describe the information that must be
represented along with the form (e.g. timestamp format) and its
necessity (e.g. required or optional). Certain fields are expressed as
references to other fields (e.g.
\textquotesingle Controller\textquotesingle{} in
\textquotesingle Processing\textquotesingle{} section is a reference to
an instance or record in \textquotesingle Parties\textquotesingle{}
section).

The Consent Receipt in ISO-27560 contains only two required
fields representing a unique identifier for the receipt and the schema
version used for the structuring of information. The information and
contents are undefined and left to each implementor to specify. A
receipt can optionally contain the entirety of the information within a
consent record or can also contain multiple consent records or other
information not within a particular consent record. Similarly, a receipt
can be made to contain only references to information within a record
without containing the information itself e.g. providing only the
consent record identifier without the contents of the record itself.

Considering the practical application of consent receipts require them
to provide information to data subjects, for the implementation
described in this document, it is assumed that the consent receipt
provides all information contained within a consent record i.e. a
receipt is a copy of the record provided to another entity. This is in
line with ISO-27560 guidance which states that the receipt
may contain the same fields as that of a consent record, and that the
mandatory fields in a consent record are also mandatory in a consent
receipt. Further, ISO-27560 allows creating different
\textquotesingle schemas\textquotesingle{} (which we call
\textquotesingle profiles\textquotesingle) to indicate changes in
requirements and their interpretations, 
through which we provide profiles for our defined implementations.

\section{Comparing ISO-27560, ISO-29184, and GDPR}\label{sec:comparison}
ISO-27560 uses prior terminology established in ISO standards, primarily defined in ISO/IEC 29100:2011 Privacy Framework \cite{ISO29100}. To support readers unfamiliar with the ISO terminology, \cref{table:mapping-29100-gdpr} provides a mapping between ISO-29100 and GDPR
terminology regarding the fundamental concepts associated with personal data
processing. Note that this mapping only provides relevant concepts and does not indicate that the concepts are interpreted in the exact same way - for example Sensitive PII in ISO terminology is \textit{similar} to Special Category personal data under GDPR, but they cannot be used interchangeably. Therefore, when applying ISO standards to GDPR, such mappings are indicative of which concepts should be (re-)interpreted with GDPR's definitions and requirements.

\begin{table}[ht]
    \centering
    \caption{Mapping between ISO/IEC 29100:2011 and EU GDPR terminology}\label{table:mapping-29100-gdpr}
    \begin{tabular}{|p{5cm}|p{7cm}|}
    \hline
        \textbf{ISO/IEC 29100:2011} & \textbf{EU GDPR} \\ \hline
        Section 2.4 Consent & Article 4-11 Consent \\ \hline
        Section 2.9 PII & Article 4-1 Personal Data \\ \hline
        Section 2.10 PII Controller & Article 4-7 Controller \\ \hline
        Section 2.11 PII Principal & Article 4-1 Data Subject \\ \hline
        Section 2.12 PII Processor & Article 4-8 Processor \\ \hline
        Section 2.23 Processing of PII & Article 4-2 Processing \\ \hline
        Section 2.26 Sensitive PII & Article 9 Special Categories of Personal Data \\ \hline
        Section 2.27 Third Party & Article 4-10 Third Party \\ \hline
    \end{tabular}
\end{table}

In prior work \cite{pandit2021comparison}, we analysed and compared ISO-29184 requirements for notice and consent with those in GDPR to understand the extent to which ISO-29184 standard can be applied to demonstrate compliance with the requirements of the GDPR. We also explored the possibility of using ISO-29184 certifications under GDPR for consent and notice. In continuation of that work, \cref{table:mapping-27560-29184-gdpr} compares ISO-27560 for consent information and ISO-29184 for privacy notice information with the requirements under GDPR to provide a holistic view of how the two standards can be used to address GDPR's requirements. In this, it is important to note that unlike ISO-29184 which is an international standard, ISO-27560 is what ISO terms a \textit{Technical Specification (TS)} which only provides guidance and is intended to obtain feedback to create a (future) international standard.

\begin{longtable}{|p{3cm}|p{7cm}|p{7cm}|}
    \caption{Mapping information requirements across ISO/IEC TS 27560:2023, ISO/IEC 29184:2020 and EU GDPR. For GDPR, numbers without prefixes are Articles, and with prefix R are Recitals} \label{table:mapping-27560-29184-gdpr} \\
    \hline
        ISO/IEC TS 27560:2023 & ISO/IEC 29184:2020 & EU GDPR \\ \hline
        3.1 consent & - & 4-11 definition of consent \\ \hline
        3.2 consent receipt & Annex B & R42, 7-1 demonstrating consent \\ \hline
        3.3 consent record & - & R42, 7-1, 13, 14, 30 recording information related to consent \\ \hline
        3.4 consent type & 5.4.3 Informed and freely given consent. 3.1 explicit consent & R32, R43, 6-1a, 9-2a conditions for consent. R42 demonstrating consent. 8 child’s consent. 9-2a explicit consent \\ \hline
        6.2 recordkeeping for privacy notices and consent & - & R42, 7-1, 13, 14, 30 recording and demonstrating consent \\ \hline
        6.2.2.1 presentation of notice & 5.2.2 providing notice, 5.2.3 appropriate expression, 5.2.7 appropriate forms, 5.2.9 accessibility & R32, R42, R43, R58, 7-2 notice for consent \\ \hline
        6.2.2.2 timeliness of notice & 5.2.5 appropriate timing & R32, R42, R43, R60, 7-2, 13, 14 notice for providing information and requesting consent. R61, 13-2 14-3 timing of notice. R62 exceptions \\ \hline
        6.2.2.3 obtaining consent & 5.2.7 appropriate forms & R42, 7-1 record of consent \\ \hline
        6.2.2.4 time and manner of consent & 5.2.6 appropriate locations & R32, R42, R43, 7-2 \\ \hline
        6.2.2.5 technical implementation & - & R42, 7-1, 13, 14, 30 maintaining information for demonstrating consent \\ \hline
        6.2.2.6 unique reference & 5.2.8 ongoing reference & R42, 7-1 demonstrating consent \\ \hline
        6.2.2.7 legal compliance & - & R39, 5 principles, 5 principles. R40, R41, 6 lawfulness and legal basis. R50 further processing. R42, 7-1 record of consent \\ \hline
        6.3.4.1 privacy\_notice & 3.2 notice & R32, R42, R43, R60, R61, 7-2, 13, 14 notice for providing information \\ \hline
        6.3.4.2 language & 5.2.4 multi-lingual notice & R32, R42, R43 conditions for consent \\ \hline
        6.3.4.3, 6.3.4.4 purposes & 5.3.2 purpose description, 5.3.3 Presentation of purpose description & 5, 6-1a, 13-1c, 13-3, 14-1c, 14-4, 15-a, 30-1b purpose of processing \\ \hline
        6.3.4.6 lawful\_basis & 5.3.15 Basis for processing & R40, R41, 6-1a, 7-1a, 9-2a, 13-1c, 13-1d, 14-1c lawfulness and legal basis \\ \hline
        6.3.4.7 pii\_information & 3.3 element of PII & 4-1, 14-1d, 15-1b, 30-1c personal data \\ \hline
        6.3.4.8 pii\_controllers & 5.3.4 Identification of the PII controller & 13-1a, 14-1a, 30-1a identity of controller \\ \hline
        6.3.4.9 collection\_method & 5.3.5 PII collection, 5.3.6 Collection method, 5.3.7 Timing and location of the PII collection & R61, 13-1, 14-1, 14-2f, 15-1g source of personal data \\ \hline
        6.3.4.10 processing\_method & 5.3.8 Method of use & 4-2, 30-2b processing methods. 13-2f, 14-2g, 15-1h automated decision making and profiling \\ \hline
        6.3.4.11 storage\_locations & 5.3.9 Geo-location of, and legal jurisdiction over, stored PII & 13-1f, 14-1f, 15-2 storage or processing location \\ \hline
        6.3.4.12 retention\_period & 5.3.11 Retention period & 13-2a, 14-2a, 15-1d, 30-1f storage duration or time limits \\ \hline
        6.3.4.13 processing\_locations & 5.3.9 Geo-location of, and legal jurisdiction over, stored PII & 13-1f, 14-1f, 15-2 processing location (including data transfers) \\ \hline
        6.3.4.14 geographic\_restrictions & 5.3.9 Geo-location of, and legal jurisdiction over, stored PII & 13-1f, 14-1f, 15-2, 30-1e, 30-2c, 44, 45, 46, 47, 48, 49-1a geographic condition (e.g. third country) \\ \hline
        6.3.4.16 jurisdiction & 5.3.9 Geo-location of, and legal jurisdiction over, stored PII & 13-1f, 14-1f, 15-2, 30-1e, 30-2c, 44, 45, 46, 47, 48, 49-1a geographic condition (e.g. third country) \\ \hline
        6.3.4.17 recipient\_third\_parties & 5.3.10 Third-party transfer & 4-9, 4-10, 13-1e, 14-1e, 15-1c, 19, 30-1d recipients \\ \hline
        6.3.4.18 withdrawal\_method & 5.3.12 Participation of PII principal & R42, 7-3, 13-2c, 14-2d withdrawing consent \\ \hline
        6.3.4.19 privacy\_rights & 5.3.12 Participation of PII principal & 13-2b, 13-2c, 14-2c, 14-2d, 15-1e, 16, 17, 18, 20, 21, 22 rights of data subject \\ \hline
        6.3.4.20 codes\_of\_conduct & - & 24-3, 32-3, 35-8, 40 codes of conduct, 42 certification \\ \hline
        6.3.4.21 impact\_assessment & 5.3.16 Risks & R75, R84 risks and risk evaluation. R90, R91, R92, R93, 35 Data Protection Impact Assessments (DPIA) \\ \hline
        6.3.4.22 authority\_party & 5.3.13 Inquiry and complaint & 13-2d, 14-2e, 15-1f complaint to authority. 36-1 consult with authority for impact assessment. 51 supervisory authority, 56 lead authority.  \\ \hline
        6.3.5.1 pii\_type & 3.3 element of PII & 4-1, 14-1d, 15-1b, 30-1c personal data types and categories \\ \hline
        6.3.5.2 pii\_attribute\_id & 3.3 element of PII & - \\ \hline
        6.3.5.3 pii\_optional & 5.4.6 Separate consent to necessary and optional elements of PII & R90, R91, 5, 13-2e, 35 optionality or necessity of personal data \\ \hline
        6.3.5.4 sensitive\_pii\_category & - & R51 protecting sensitive data \\ \hline
        6.3.5.5 special\_pii\_category & - & R51, R53, R71, 6, 9, 22-4, 30-1c, 35 special categories of personal data \\ \hline
        6.3.6.6 party\_name & - & 13-1a, 14-1a, 30-1a, 30-2a \\ \hline
        6.3.6.7 party\_role & - & 4-1, 4-7, 4-7, 4-8, 4-9, 4-10, 13-1a, 13-1e, 14-1a, 14-1e, 26-1, 28, 30-1a, 30-1d, 30-2a, 37 \\ \hline
        6.3.6.8 party\_contact & - & 13-1a, 13-1b, 14-1a, 14-1b, 26-1 \\ \hline
        6.3.6.9 party\_type & - & 4-1, 4-7, 4-7, 4-8, 4-9, 4-10, 13-1e, 14-1e, 15-1c, 19 \\ \hline
        6.3.7.1 event\_time & 5.4.8 Timeliness & R42, 7-1, 13, 14, 30 maintaining information for demonstrating consent \\ \hline
        6.3.7.2 validity\_duration & 5.4.7 Frequency & 25 Data Protection by Design and by Default \\ \hline
        6.3.7.3 entity\_id & - & R42, 4-11, 6-1a, 7-3, 8-1, 8-2, A13, A14 \\ \hline
        6.3.7.4 event\_type & 5.4.3 Informed and freely given consent & 4-11 (regular) consent. 9-2a explicit consent. R32, 7-1 given consent. R32, 7-2 request for consent \\ \hline
        6.3.7.6 event\_state & 5.5.2 Renewing notice, 5.5 Change of conditions, 5.5.3 Renewing consent & 4-11, 6-1a, 9-2a given consent. R42, 7-3 withdrawn consent. \\ \hline
        6.4.3 consent management & - & R32, R42, R43, R60, R61, 7-2, 12, 13, 14 information about given consent and applicable rights, R42, 7-3 withdrawing consent \\ \hline
        6.4.4 PII principal participation & 5.3.12 Participation of PII principal, 5.3.14 Information about accessing the choices made for consent & R32, R42, R43, R60, R61, 7-2, 12, 13, 14 information about given consent and applicable rights, R42, 7-3 withdrawing consent \\ \hline
        6.4.6 receipt content & 5.3.14 Information about accessing the choices made for consent & R32, R42, R43, R60, R61, 7-2, 12, 13, 14 information about given consent and applicable rights \\ \hline
        Annex B consent lifecycle & 5.5.2 Renewing notice, 5.5 Change of conditions, 5.5.3 Renewing consent & 4-11, 6-1a, 9-2a given consent. R42, 7-3 withdrawn consent. \\ \hline
        Annex E security of consent records and receipts & - & R75, R76, R77, R78, R83, 24, 25, 30, 32, 44 \\ \hline
        Annex F signals communicating PII Principal’s preferences and decisions & - & R32, 7-2, 21-5 \\ \hline
\end{longtable}

\section{Consent Records and Receipts using DPV}\label{sec:dpv}
ISO-27560 only defines the information fields and does not prescribe how they should be technically represented in practice. To implement ISO-27560, the information therefore must be represented in a format such as JSON which is widely supported and easy to use. However, the use of JSON requires a strict agreement on how the information should be structured and how it should be interpreted. The JSON-LD (JSON for Linked Data) format enables use of JSON with linked data so that the ontologies and vocabularies defined using W3C standards can be exchanged as JSON data. For this reason, ISO-27560 Annex C provides examples of consent records and receipts for both JSON and JSON-LD. Implementing ISO-27560 in a machine-readable manner using JSON-LD requires agreement on the schema or ontology to represent the fields. The Annex C JSON-LD example uses the Data Privacy Vocabulary\footnote{\url{https://w3id.org/dpv}} (DPV) \cite{panditCreatingVocabularyData2019,panditDataPrivacyVocabulary2024} which is maintained by the W3C Data Privacy Vocabularies and Controls Community Group\footnote{\url{https://www.w3.org/groups/cg/dpvcg/}} (DPVCG). 

DPV is a state of the art resource that provides the necessary ontology to represent concepts such as purpose, processing operations, personal data, legal roles, as well as a rich and extensive taxonomy expanding on each of these concepts to enable representing of practical use-cases. For example, using DPV, it is possible to exchange ISO-27560 records and receipts in JSON-LD which specify the \textit{Purpose} is \textit{Marketing} in an interoperable manner. In addition, DPV also features extensions through which different jurisdiction specific concepts can be represented, and for which it provides extensions for EU regulations such as the GDPR, DGA, and the upcoming AI Act. This enables flexibility of expression general requirements such as the \textit{legal basis} should be \textit{consent}, as well as specific requirements such as \textit{explicit consent} as per GDPR Art.9-1a. 

DPV was initiated as part of the SPECIAL H2020 project and has been developed for over 6 years with a multi-disciplinary community made up of computer scientists, legal experts, sociologists, data protection officers, industry stakeholders, and authorities. It has been actively used in several projects at national and international (e.g. Horizon Europe) levels, is being used by the industry, and has been acknowledged by within standards (including ISO-27560) \cite{panditDataPrivacyVocabulary2024}. As such, it represents the best resource currently available for representing consent records and receipts as well as other legally relevant information in a machine-readable form that is based on open (free and non-proprietary) interoperable standards.

To support the implementation of use of DPV in implementing consent records and receipts in conformance with ISO-27560 and the GDPR, we have developed a technical specification which can be accessed online at \url{https://w3id.org/dpv/guides/consent-27560}. The specification describes how the required fields in ISO-27560 and GDPR are represented using DPV (summarised below) and provides illustrative examples for each (see online). Consent records are represented as instances of the concept \texttt{dpv:ConsentRecord} and receipts are represented as instances of the concept \texttt{dpv:ConsentReceipt}. For reviewers convenience: complete examples of a consent record and a consent receipt are provided in the annexes at the end of this article.

\textbf{Profiles:} To support implementing ISO-27560 as well as its use to comply with GDPR, DPVCG defines 4 schemas or profiles defined under the
namespace \url{https://w3id.org/dpv/schema/dpv-27560\#} (prefixed as \texttt{dpv-27560:}).

\begin{enumerate}
\item
  \href{https://w3id.org/dpv/schema/dpv-27560\#record}{\textbf{\texttt{dpv-27560:record}}}:
  Consent Records conforming with ISO-27560.
\item
  \href{https://w3id.org/dpv/schema/dpv-27560\#record-eu-gdpr}{\textbf{\texttt{dpv-27560:record-eu-gdpr}}}
  Consent Records conforming with ISO-27560 and containing information as
  required by EU GDPR.
\item
  \href{https://w3id.org/dpv/schema/dpv-27560\#receipt}{\textbf{\texttt{dpv-27560:receipt-record}}}
  Consent Receipts conforming with ISO-27560 and providing information from consent record(s).
\item
  \href{https://w3id.org/dpv/schema/dpv-27560\#receipt-eu-gdpr}{\textbf{\texttt{dpv-27560:receipt-eu-gdpr}}}
  Consent Receipts conforming with ISO-27560 and providing information from consent record(s) as required by EU GDPR.
\end{enumerate}

\textbf{Metadata Fields:} (see table \ref{table:metadata}) to describe the generic metadata fields associated with records and receipts, we utilise the DCMI Metadata Terms standard\footnote{\url{https://dublincore.org/specifications/dublin-core/dcmi-terms/}} (prefixed as \texttt{dct:}). A consent record or receipt indicates use of the DPV profiles by using \texttt{dct:conformsTo} with one of the profiles described above.

\begin{table}[!ht]
    \centering \label{table:metadata}
    \caption{DPV concepts for ISO/IEC 27560:2023 Metadata fields} \label{table:dpv-27560-metadata}
    \begin{tabular}{|p{3cm}|p{2cm}|p{2.5cm}|p{3.5cm}|}
    \hline
        \textbf{Field} & \textbf{Cardinality} & \textbf{DPV Concept} & \textbf{DPV Property} \\ \hline
        Schema Version & 1 & N/A & 	exttt{dct:conformsTo} \\ \hline
        Record Identifier & 1..* & N/A & \texttt{dpv:hasIdentifier} \\ \hline
        Data Subject & 1 & \texttt{dpv:DataSubject} & \texttt{dpv:hasDataSubject} \\ \hline
    \end{tabular}
\end{table}

\textbf{Processing Fields:} (see table \ref{table:processing}) ISO-27560 contains 22 fields related to processing activities, and 5 additional fields regarding personal data involved in processing. The structuring of these fields within ISO-27560 is of the form where the "PII Processing" section contains an array of "purposes" where each "purpose" is expressed with its own fields regarding legal basis, collection method, storage locations, and so on. Within the DPV implementation, this is replaced with \texttt{dpv:Process} where each 'process' represents a distinct processing activity with its own fields e.g. purposes, personal data, recipients. Thus a consent record and receipt may cover multiple processes (and purposes) which permits an unambiguous and exact representation e.g. which purpose, implemented by which entity, with what data, recipient, etc.

\begin{longtable}{|p{4cm}|p{2cm}|p{5.7cm}|p{5.5cm}|}
    \caption{DPV concepts for ISO/IEC 27560:2023 Processing fields} \label{table:processing}\\
    \hline
        \textbf{Field} & \textbf{Cardinality} & \textbf{DPV Concept} & \textbf{DPV Property} \\ \hline
        Process & 1..* & \texttt{dpv:Process} & \texttt{dpv:hasProcess} \\ \hline
        Purpose & 1..* & \texttt{dpv:Purpose} & \texttt{dpv:hasPurpose} \\ \hline
        Personal Data & 1..* & dpv: & \texttt{dpv:hasPersonalData} \\ \hline
        Personal Data Type & 1..* & \texttt{dpv:PersonalData} taxonomy & \texttt{dpv:hasPersonalData} or \texttt{dct:type} \\ \hline
        Personal Data Identifier & 0..* & N/A & dct:identifier \\ \hline
        Personal Data Necessity & 0..* & \texttt{dpv:Necessity} & \texttt{dpv:hasNecessity} \\ \hline
        Sensitive/Special Category & 0..* & \texttt{dpv:SensitivePersonalData}, \texttt{dpv:SpecialCategoryPersonalData} & \texttt{dpv:hasPersonalData} or \texttt{dct:type} \\ \hline
        Processing Operations & 0..* & \texttt{dpv:Processing} & \texttt{dpv:hasProcessing} \\ \hline
        Data Source & 0..* & \texttt{dpv:DataSource} & \texttt{dpv:hasDataSource} \\ \hline
        Storage Condition & 1..* & \texttt{dpv:StorageCondition}, \texttt{dpv:StorageLocation}, \texttt{dpv:StorageDuration}, \texttt{dpv:StorageDeletion} & \texttt{dpv:hasStorageCondition} \\ \hline
        Processing Condition & 0..* & \texttt{dpv:ProcessingCondition}, \texttt{dpv:ProcessingLocation}, \texttt{dpv:ProcessingDuration} & \texttt{dpv:hasProcessingCondition} \\ \hline
        Geographic Restriction & 0..* & \texttt{dpv:Rule} & \texttt{dpv:hasRule} \\ \hline
        Data Controller & 1..* & \texttt{dpv:DataController} & \texttt{dpv:hasDataController} \\ \hline
        Legal Basis & 0..* & \texttt{dpv:LegalBasis} & \texttt{dpv:hasLegalBasis} \\ \hline
        Recipients & 1..* & \texttt{dpv:Recipient} & \texttt{dpv:hasRecipient} \\ \hline
        Consent Change \&amp; Withdrawal & 1..* & \texttt{dpv:InvolvementControl}, \texttt{dpv:WithdrawingFromActivity} & \texttt{dpv:hasInvolvementControl} \\ \hline
        Jurisdiction & 1..* & \texttt{dpv:Jurisdiction} & \texttt{dpv:hasJurisdiction} \\ \hline
        Rights & 1..* & \texttt{dpv:DataSubjectRight} & \texttt{dpv:hasRight} \\ \hline
        Services & 0..* & \texttt{dpv:Service} & \texttt{dpv:hasService} \\ \hline
        Code of Conduct & 0..* & \texttt{dpv:CodeOfConduct} & \texttt{dpv:hasOrganisationalMeasure} \\ \hline
        Impact Assessment & 0..* & \texttt{dpv:ImpactAssessment} & \texttt{dpv:hasAssessment} \\ \hline
        Notice & 1..* & \texttt{dpv:Notice} & \texttt{dpv:hasNotice} \\ \hline
        Notice Language & 1..* & N/A & dct:language \\ \hline
    
\end{longtable}

\textbf{Entity Fields:} (see table \ref{table:entity}) DPV uses the term \textit{Entity} for what ISO-27560 refers to as \textit{Party}. Entities are expressed using instances of \texttt{dpv:Entity} and associated using \texttt{dpv:hasEntity}. DPV also distinguishes between \textit{Entities} and \textit{Legal Entities} - and their representatives or agents, through which it can be accurately represented whether a party in a consent record acted on their own or it was someone acting on their behalf. This is of relevance for implementations such as consent for children which involves parents or guardians, or even data intermediaries under DGA which can act to support individuals in consent decision making.

\begin{table}[!ht]
    \centering \label{table:entity}
    \caption{DPV concepts for ISO/IEC 27560:2023 Party fields} \label{table:dpv-27560-parties}
    \begin{tabular}{|p{3cm}|p{2cm}|p{4cm}|p{4cm}|}
    \hline
        \textbf{Field} & \textbf{Cardinality} & \textbf{DPV Concept} & \textbf{DPV Property} \\ \hline
        Name & 1..* & N/A & \texttt{dpv:hasName} \\ \hline
        Identifier & 1..* & N/A & \texttt{dpv:hasIdentifier} \\ \hline
        Role & 1..* & \texttt{dpv:DataController}, \texttt{dpv:DataProcessor}, \texttt{dpv:ThirdParty}, \texttt{dpv:Authority}, \texttt{dpv:DataSubject} & \texttt{dpv:hasEntity}, \texttt{dpv:hasDataController}, \texttt{dpv:hasDataProcessor}, \texttt{dpv:hasThirdParty}, \texttt{dpv:hasAuthority}, \texttt{dpv:hasDataSubject} \\ \hline
        Contact & 1..* & \texttt{schema:ContactPoint} & \texttt{schema:contactPoint} \\ \hline
        Postal Address & 1..* & \texttt{schema:PostalAddress} & \texttt{schema:contactPoint} \\ \hline
        Email & 0..* & N/A & \texttt{schema:email} \\ \hline
        Phone & 0..* & N/A & \texttt{schema:telephone} \\ \hline
        URL & 0..* & N/A & \texttt{schema:url} \\ \hline
    \end{tabular}
\end{table}

\textbf{Consent Event Fields:} (see table \ref{table:events}) These fields are used to indicate the type of consent (e.g. \textit{Implicit}, \textit{Expressed}, \textit{Explicit}) as expressed by the data subject. In DPV, \texttt{dpv:ConsentType} represents consent types to be used as a legal basis and has the following different types: \textit{Informed} with specialisations for \textit{Implied} when implied or given by an indirect action (e.g. merely browsing a website), \textit{Expressed} for  direct expressed action (e.g. a checkbox), and \textit{ExplicitlyExpressed} for direct action concerning solely the consent in context. To indicate consent types as defined in GDPR, the DPV's GDPR extension is used e.g. \texttt{eu-gdpr:A6-1-a} for expressed consent and \texttt{eu-gdpr:A9-2-a} for explicit consent. 

In addition to the type of consent, these fields also enable expressing the status of consent, such as whether it has been requested, given, refused, expired, terminated, invalidated, or re-affirmed. Each event can contain metadata to indicate when it took place (e.g. date when consent was given), how it was expressed (e.g. in the account dashboard), its duration (e.g. validity of given consent), and by whom (e.g the data subject).

\begin{table}[!ht]
    \centering \label{table:events}
    \caption{DPV concepts for ISO/IEC 27560:2023 Event fields} \label{table:dpv-27560-events}
    \begin{tabular}{|p{3cm}|p{2cm}|p{4cm}|p{4cm}|}
    \hline
        Field & Cardinality & DPV Concept & DPV Property \\ \hline
        Consent Type & 1..* & \texttt{dpv:Consent} taxonomy & \texttt{dpv:hasLegalBasis} \\ \hline
        Consent State & 1..* & \texttt{dpv:ConsentStatus} taxonomy & \texttt{dpv:hasConsentStatus} \\ \hline
        Event Time & 1..* & N/A & \texttt{dpv:isIndicatedAtTime} \\ \hline
        Event Duration & 1..* & \texttt{dpv:Duration} & \texttt{dpv:hasDuration} \\ \hline
        Expression by Entity & 1..* & \texttt{dpv:Entity} & \texttt{dpv:isIndicatedBy} \\ \hline
        Expression Method & 0..* & N/A & \texttt{dpv:hasIndicationMethod} \\ \hline
    \end{tabular}
\end{table}

\textbf{Consent Receipts:} (see table \ref{table:receipts}) ISO-27560 only defines the schema version and receipt identifier fields for consent receipts. For other fields, it recommends using the same fields as that of a consent record. In its guidance, it states that the mandatory fields in consent records should also be mandatory in receipts. Based on this, we only define the additional fields for consent receipts and suggest reusing the consent record fields with their necessity/optionality requirements. Therefore, a consent receipt only has three mandatory fields with the rest of the information being present as instances of consent record(s).

\begin{table}[!ht]
    \centering \label{table:receipts}
    \caption{DPV concepts for ISO/IEC 27560:2023 Receipt Metadata fields} \label{table:dpv-27560-processing}
    \begin{tabular}{|l|l|l|l|}
    \hline
        \textbf{Field} & \textbf{Cardinality} & \textbf{DPV Concept} & \textbf{DPV Property} \\ \hline
        Schema Version & 1 & N/A & \texttt{dct:conformsTo} \\ \hline
        Receipt Identifier & 1..* & N/A & \texttt{dpv:hasIdentifier} \\ \hline
        Consent Record & 1..* & \texttt{dpv:ConsentRecord} & \texttt{dpv:hasRecordOfActivity} \\ \hline
    \end{tabular}
\end{table}

\section{Supporting GDPR and DGA}\label{sec:gdpr-dga}

\textbf{Using ISO-27560 and ISO-29184 within the EU legal framework:} ISO-27560 and ISO-29184 are developed and governed by the International Standards Organisation (ISO), and are not specific to EU's regulations and terminology. To support their use in the legal frameworks, they need to be approved as `Euronorm' (EN) through an EU standardisation body such as CEN, CENELEC, or ESO. At the moment, ISO-29184 has already been approved as EN, and we are working on a proposal with the Irish and Swedish national bodies to recommend the adoption of ISO-27560 as EN. Further, we have also submitted a proposal to the relevant ISO committees to make ISO-27560 standard freely accessible as its guidance is valuable for responsible innovation.

Having these standards as EN provides a strong framework for their utilisation in regulations, such as for notice and consent under GDPR. However, merely adopting the standards on an `as-is' basis will not be sufficient. For example, the terminology in 29184 and GDPR has crucial differences which must be identified and appropriate guidance developed to enable using ISO-29184 with GDPR \cite{pandit2021comparison}. Similarly, to address current issues regarding consent \cite{mattePurposesIABEurope2020,machuletzMultiplePurposesMultiple2020} and further studies are required to assess the extent of these standards in solving existing issues and what additional measures need to be adopted beyond conformance with the standards.

\textbf{Demonstrating consent under GDPR:} GDPR Article 7-1 creates an obligation for data controllers to maintain consent information and to keep it up to date with the goal of demonstrating where consent was given, refused, or withdrawn. ISO-27560 provides a standard for a common technical structure to support implementing this obligation. In addition to this, GDPR Article 13 and Article 14, amongst others, also require record keeping for what information was provided to individuals in order to implement informed consent. ISO-29184 provides a standard for describing privacy notices, and together with ISO-27560 enables maintaining records of what information was provided and the resulting consent decisions. Based on the analysis provided in this article that demonstrates applicability of ISO-27560 and ISO-29184 to GDPR, we recommend authorities to suggest using these standards to support GDPR compliance.

\textbf{Receipts to support rights under GDPR:} ISO-27560 contains fields for acknowledging which rights exist, and with DPV we can express how/where to exercise them and what information will be required (e.g. identity verification). Further, consent decisions (e.g. given, withdrawn) are themselves also personal data about the data subject, and therefore subject to rights such as Art.20 data portability. This can be a way to enable the use of receipts under GDPR even where it is not explicitly defined as a concept by considering consent information as \textit{personal data}. Considering consent information as personal data makes it subject to the right to data portability under Article 20 which requires providing information ``in a structured, commonly used and machine-readable format''. Further, Article 20 also allows `` the right to transmit those data to another controller'', which can be utilised to transfer consent decisions from one controller to another - a crucial mechanism for the implementation of data reuse and altruism under DGA.

\textbf{Common consent form under DGA:} Article 25 of the DGA requires the Commission to produce a common consent form that will provide information in both human- and machine-readable forms. ISO-27560 with ISO-29184, based on the analysis in this article demonstrating their usefulness to meet GDPR requirements, should be used to define what information should be present in these forms. ISO-29184 as the standard for privacy notices provides the human-oriented representation of information in the consent form, and ISO-27560 and the DPV implementation provide the machine-readable representation. The advantage of using these standards is that the resulting solution would be useful not only in EU but globally due to the global scope of ISO. The advantage of using DPV here is in providing common semantics based on W3C standards that support extensions for specific jurisdictions (like EU with GDPR and DGA) and its extensive taxonomy supporting practical use-cases which promote interoperability. \textbf{Through direct meetings, we have presented this work to the EU Commission's Unit G.1 which looks after GDPR and DGA implementations.}

\textbf{Data Intermediaries under DGA:} We are also working on further implementations to support DGA by developing specific technical specifications that define how data intermediaries should maintain consent records and issue receipts, and support them in their duties by providing a way to express data reuse requests in a machine-readable form that can be matched with the consent to ensure the purposes are compatible in accordance with the GDPR. This will be based on existing work \cite{estevesODRLProfileExpressing2021} that utilises the W3C Open Digital Rights Language (ODRL) standard \cite{iannellaODRLInformationModel2018} for representing policies and agreements, and using it in combination with DPV to create DGA specific \textit{offers} for data subjects and data intermediaries to indicate which data is available for reuse and under what conditions, {requests} for data users to indicate what data they are looking for, and \textit{agreements} to represent the conditions under which data reuse has been approved. We have already demonstrated the feasibility of using ODRL and DPV for such an approach in a manner that improves both technical and organisational processes for the use-case of sharing genomic health datasets \cite{panditEnhancingDataUse2024}. 

\textbf{Data Reuse and Altruism under DGA:}
To support the DGA's goals of reusing data for altruism, we are working on creating a taxonomy of altruistic purposes within DPV and developing a framework to express them in a manner that is compatible with GDPR's requirements for consent and information keeping based on ISO-27560. We are also working on novel approaches such as assessing the compatibility of ISO-27560 defined consent records with information required in a Data Protection Impact Assessment (DPIA), through which we aim to enable data subjects or data intermediaries to conduct their own DPIAs based on a common registry of risks and mitigations provided through the DPV. Through this we aim to establish responsible  practices while promoting data reuse and altruism.

\section{Implementation Considerations and Future Work}\label{sec:considerations}
\subsection{Trust and Security}\label{sec:considerations-security}
Security considerations are extremely important in the implementation of consent records and receipts, with ISO-27560 Annex E providing guidance for implementations. Consent records are intended to be maintained internally by an entity, and require measures to ensure they maintain their consistency and correctness, and are not tampered with. This includes best practices for information management such as using cryptographic hashes to ensure information has not changed, or using access control to ensure only authorised modifications are permitted. Current internationals standards such as W3C Decentralized Identifiers\footnote{\url{https://www.w3.org/TR/vc-data-model/}} (DID) and W3C Verifiable Credentials\footnote{\url{https://www.w3.org/TR/did-core/}} (VC) allow for implementations compatible with the implementation of ISO-27560 using DPV as all are based on interoperable semantic web standards.

For consent receipts to be utilised in a verifiable and trustworthy manner, the information provided within the receipt may require cryptographic measures to provide assurance to prove its immutability and non-repudiation. Further, receipts are intended to be information provided or exchanged between different entities, which may necessitate a mechanism to demonstrably verify the provenance (e.g. a receipt was provided by A to B) and its immutability (e.g. receipt contained X exactly). Cryptography techniques such as digital signatures and certificates can support such applications based on their current utilisation in internet-enabled applications and documentations. Prior work \cite{jesusConsentReceiptsUsable2022} and projects\footnote{NGI funded Privacy as Expected: Consent Gateway project D2 Final Technical Deliverable \url{https://doi.org/10.5281/zenodo.5086238}} have explored such considerations, but effective implementation requires consensus amongst stakeholders to create an interoperable ecosystem.

Given the role of consent records and receipts in demonstrating consent decisions, they may end up with potentially sensitive information. ISO-27560 recommends not putting such information directly in records and receipts, and if necessary then implementations should utilise techniques such as information masking or pseudonymisation to avoid directly exposing sensitive information. - though this has to be balanced with the purpose of receipts in providing data subjects with information about their consent.

\subsection{Using Records and Receipts with eIDAS and EUDI Wallet}\label{sec:considerations-wallets}
Following the launch of projects for using European Digital Identity wallet (EUDI) wallet\footnote{\url{https://digital-strategy.ec.europa.eu/en/news/eu-digital-identity-4-projects-launched-test-eudi-wallet}} for travel, health, banking, education and other sectors,
CEN TC224 WG20\footnote{\url{https://www.cencenelec.eu/areas-of-work/cen-sectors/digital-society-cen/information-and-identification-systems/}}, which is the EU standardisation body's technical committee for personal identification, has initiated a new standards project to provide guidance on when personal data (attributes) are shared from the wallet in compliance with eIDAS and its proposed revision. 

In this, ISO-27560 and ISO-29184 can be used to create an interoperable and standards based mechanism to structure information and ensure the mandatory fields needed to comply with GDPR are present. Further, the use of these standards also enables a consistent approach for creating common privacy dashboards that can work across EU. Such privacy dashboards would allow a wallet holder to have an overview of all their consent transactions, including any pending requests as well as provide a centralised mechanism for controlling their rights and withdrawing consent by using the eIDAS and eID mechanisms to establish identity and proof of past engagement.

ISO-27560 and ISO-29184 are also crucial as being the only standards regarding consent records and receipts, and privacy notices respectively. Using the analysis and implementations described in this article, a ISO-27560 solution that is also conformant with the GDPR can be used to store consent records and receipts in wallets, which enables data subjects to have a copy of their decision and agreement to process personal data. 

Having this information made available to the data subject in a machine-readable format further enables its use in innovative applications that promote reuse of data while ensuring adequate adherence to the EU's values and regulations. For example, by looking at past consent records or receipts, preferences can be identified for how the individual makes decisions and these can be used to create a template or pattern that will make future consent decisions more efficient and simpler for the individual. ISO-27560 Annex F provides guidance on how such preferences used as 'privacy signals' can be represented within consent records and receipts.

Another powerful paradigm is also made possible when combining ISO-27560 with eID, eIDAS, and EUDI - where the data subject initiates the consent process by providing a specific consent to use or reuse their personal data, for example to access a particular service. In this scenario, the data subject decides the extent and limit of what their consent will cover, provides their consent to the service provider, and maintains a consent record within their wallet with a signed receipt provided to the service provider as proof of consent.

\subsection{Standard for PII Processing Record Information}
Even though ISO-27560 only focuses on consent records and receipts, its fields were developed with the intention of a future expansion in a separate standard to cover other legal bases, such as the 7 other legal basis in GDPR Article 6. To continue in this direction, we have initiated a `new standard' proposal in ISO regarding `PII processing record information'. 

To support this activity, we are currently identifying the specific requirements for record keeping for each legal basis and creating the necessary specifications using DPV. This builds on prior work providing a machine-readable Records of Processing Activities (ROPA) required under GDPR Article 30, and which consolidates the guidelines from all 30 EU/EEA member states. 

\subsection{Technical Considerations in Managing Records and Receipts}
We can use the Data Catalog Vocabulary\footnote{DCAT - Version 3 \url{https://www.w3.org/TR/vocab-dcat-3/}} (DCAT), a W3C standard, to represent the records as datasets and receipts as a catalogue of records. By doing so, the metadata fields provided by DCAT can be readily used to represent information that supports in maintenance and exchange of consent records and receipts, including using existing infrastructure to manage them. DCAT is a widely used standard that supports implementing (open) data portals and has tooling for discovery and management of information. The EU has developed the DCAT Application Profile\footnote{DCAT Application profile for data portals in Europe (DCAT-AP) \url{https://op.europa.eu/en/web/eu-vocabularies/dcat-ap}} (DCAT-AP) which extends DCAT to support the EU Open Data Portals\footnote{\url{https://data.europa.eu/}}. 

Through these records and receipts can be readily communicated as interoperable datasets between relevant entities - for example controller to data subject, or between controllers and third parties. This is a crucial technical enabler for the principle of increasing data value through utilisation within the Data Governance Act and Data Spaces. In particular, the use of DCAT(-AP) also supports the addition of further policies and measures to support the implementation of data intermediaries which will be required to maintain consent records under the obligations of the DGA.

\subsection{IEEE P7012 Machine-Readable Privacy Terms}
In addition to the above, we are also working with the IEEE P7012 group to develop a standard for machine-readable privacy terms which uses ISO-27560 and ISO-29184 with DPV to define the conditions under which the individual allows use or reuse of their personal data. The use of this standard will provide an efficient and optimal mechanism for data subjects to signal their consent or initiate an agreement with a service provider.

\section{Conclusion}\label{sec:conclude}
This article provided a thorough analysis of how ISO/IEC TS 27560:2023 and ISO/IEC 29184:2020 can be used to create consent records and receipts in a machine-readable format that support GDPR requirements and enable the reuse of data under the DGA. Based on this analysis, we provide a concrete argument for why these two standards should be adopted and recommended by GDPR stakeholders. We also described the ongoing efforts of the W3C Data Privacy Vocabularies and Controls Community Group (DPVCG) in creating a technical specification to support implementing ISO-27560 by using its Data Privacy Vocabulary (DPV). Our work is a significant contribution to the ongoing efforts of implementing the DGA where the Commission is required to develop a common consent form that is both human- and machine-readable. We also discussed how this work can be utilised in practice, where reported on our ongoing efforts to adopt the standard within the EU's legal framework, further develop specific implementations to support the needs of DGA, and how this work compliments the ongoing developments of eID, eIDAS2, and EUDI implementations.

\subsubsection*{Acknowledgements}
Jan Lindquist, through the Swedish National Standards Body, was a contributor and the co-editor of ISO/IEC TS 27560:2023. Harshvardhan J. Pandit, through the Irish National Standards Body, was a contributor to the ISO/IEC TS 27560:2023, and is the chair of the W3C Data Privacy Vocabularies and Controls Community Group.

This research was conducted with the financial support of Science Foundation
Ireland at ADAPT, the SFI Research Centre for AI-Driven Digital Content
Technology at Dublin City University Grant\#13/RC/2106\_P2. For the purpose of
Open Access, the author has applied a CC BY public copyright licence to any
Author Accepted Manuscript version arising from this submission.

\bibliographystyle{splncs04}
\bibliography{paper}

\appendix
\section{Example of Consent Record with both required and optional fields}

\begin{minted}{json-ld}
{
    "@id": "https://example.com/a6f58318-72e6-46a2-bfd7-f36d795e30cd",
    "@type": "dpv:ConsentRecord",
    "dct:identifier": "a6f58318-72e6-46a2-bfd7-f36d795e30cd",
    "dct:conformsTo": "https://w3id.org/dpv/schema/dpv-27560#record",
    "dpv:hasDataSubject": {
        "@id": "0760c9ba",
        "type": "dpv:Consumer"
    },
    "dpv:hasDataController": "ex:Acme",
    "dpv:hasDataProcessor": "ex:Beta",
    "dpv:hasJurisdiction": ["loc:IE"],
    "dpv:hasApplicableLaw": "eu-gdpr:GDPR",
    "dpv:hasLegalBasis": "eu-gdpr:A6-1-a",
    "dpv:hasProcess": {
        "@type": "dpv:Process",
        "dpv:hasService": "Register for Event X",
        "dpv:hasRecipient": ["ex:Acme", "ex:Beta"],
        "dpv:hasPurpose": "dpv:PaymentManagement",
        "dpv:hasPersonalData": {
          "@type": "pd:EmailAddress",
          "rdf:value": "hello@example.com",
          "dpv:hasNecessity": "dpv:Optional",
          "dpv:hasDataSource": "dpv:DataSubject",
        },
        "dpv:hasStorageCondition": [{
            "@type": "dpv:StorageLocation",
            "dpv:hasLocation": ["loc:IE", "loc:FR", "loc:DE"],
        }, {
            "@type": "dpv:StorageDuration",
            "dpv:hasDuration": "P6M",
        }, {
            "@type": "dpv:StorageDeletion",
            "dpv:hasDuration": "P1M"
        }]
    },
    "dpv:hasProcess": {
        "@type": "dpv:Process",
        "dpv:hasService": "Register for Event X",
        "dpv:hasRecipient": ["ex:Acme", "dpv:DataSubject"],
        "dpv:hasPurpose": "dpv:IdentityVerification",
        "dpv:hasPersonalData": {
          "@type": "pd:OfficialID",
          "dct:identifier": "XJ189019D",
          "dpv:hasNecessity": "dpv:Required",
          "dpv:hasDataSource": "ex:Acme",
        },
        "dpv:hasStorageCondition": [{
            "@type": "dpv:StorageLocation",
            "dpv:hasLocation": "dpv:WithinDevice",
        }, {
            "@type": "dpv:StorageDuration",
            "dpv:hasDuration": {
                "@type": "dpv:UntilEventDuration",
                "rdf:value": "Account Closure"
        }]
    },
    "dpv:hasNotice": {
        "@id": "https://example.com/notices/a6f58318-72e6-46a2-bfd7-f36d795e30cd",
        "@type": "dpv:ConsentNotice",
        "dct:date": "2024-01-01",
        "dct:language": "EN",
        "dct:coverage": "2024-01-01/P12M"
    }
    "dpv:hasImpactAssessment": {
      "@type": "dpv:DPIA",
      "schema:url": "https://example.com/DPIA"
    }
    "dpv:hasInvolvementControl": {
      "@type": ["dpv:ProvidingPermission", "dpv:WithdrawingPermission"],
      "dpv:isExercisedAt": "https://example.com/manage-consent"
    },
    "dpv:hasRight": [{
            "@type": ["dpv:DataSubjectRight", "eu-gdpr:A7-3"],
            "dct:title": "Right to Withdraw Consent",
            "dpv:isExercisedAt": "https://example.com/rights",
    },
    "dpv:hasConsentStatus": [{
        "@type": ["dpv:ConsentGiven", "dpv:ExpressedConsent"],
        "dpv:isIndicatedBy": "dpv:DataSubject",
        "dpv:hasIndicationMethod": "Interaction in App",
        "dpv:isIndicatedAtTime": "2024-01-01"
    }, {
        "@type": "dpv:ConsentWithdrawn",
        "dpv:isIndicatedBy": "dpv:DataSubject",
        "dpv:hasIndicationMethod": "Interaction in App",
        "dpv:isIndicatedAtTime": "2024-04-20"
    }]
}
\end{minted}

\section{Example of Consent Receipt with required fields from consent record}
\begin{minted}{json-ld}
{
    "@id": "https://example.com/receipt-asdmj1oasd",
    "@type": "dpv:ConsentRereceipt",
    "dct:identifier": "receipt-asdmj1oasd",
    "dct:conformsTo": "https://w3id.org/dpv/schema/dpv-27560#receipt",
    "dct:created": "2024-01-31",
    "dct:publisher": "ex:Acme",
    "schema:recipient": "dpv:DataSubject",
    "dpv:hasRecordOfActivity": {
        "@id": "https://example.com/a6f58318-72e6-46a2-bfd7-f36d795e30cd",
        "@type": "dpv:ConsentRecord",
        "dct:identifier": "a6f58318-72e6-46a2-bfd7-f36d795e30cd",
        "dct:conformsTo": "https://w3id.org/dpv/schema/dpv-27560#record",
        "dpv:hasDataSubject": {
            "@id": "0760c9ba",
            "type": "dpv:Consumer"
        },
        "dpv:hasDataController": "ex:Acme",
        "dpv:hasDataProcessor": "ex:Beta",
        "dpv:hasJurisdiction": ["loc:IE"],
        "dpv:hasApplicableLaw": "eu-gdpr:GDPR",
        "dpv:hasProcess": {
            "@type": "dpv:Process",
            "dpv:hasRecipient": ["ex:Acme", "ex:Beta"],
            "dpv:hasPurpose": "dpv:PaymentManagement",
            "dpv:hasPersonalData": "pd:EmailAddress",
            "dpv:hasStorageCondition": [{
                "@type": "dpv:StorageLocation",
                "dpv:hasLocation": ["loc:IE", "loc:FR", "loc:DE"]
            }, {
                "@type": "dpv:StorageDuration",
                "dpv:hasDuration": "P6M"
            }, {
                "@type": "dpv:StorageDeletion",
                "dpv:hasDuration": "P1M"
            }]
        },
        "dpv:hasProcess": {
            "@type": "dpv:Process",
            "dpv:hasRecipient": ["ex:Acme", "dpv:DataSubject"],
            "dpv:hasPurpose": "dpv:IdentityVerification",
            "dpv:hasPersonalData": "pd:OfficialID",
            "dpv:hasStorageCondition": [{
                "@type": "dpv:StorageLocation",
                "dpv:hasLocation": "dpv:WithinDevice"
            }, {
                "@type": "dpv:StorageDuration",
                "dpv:hasDuration": {
                    "@type": "dpv:UntilEventDuration",
                    "rdf:value": "Account Closure"
                }
            }]
        },
        "dpv:hasInvolvementControl": {
          "@type": ["dpv:ProvidingPermission", "dpv:WithdrawingPermission"],
          "dpv:isExercisedAt": "https://example.com/manage-consent"
        },
        "dpv:hasRight": {
                "@type": ["dpv:DataSubjectRight", "eu-gdpr:A7-3"],
                "dct:title": "Right to Withdraw Consent",
                "dpv:isExercisedAt": "https://example.com/rights"
        },
        "dpv:hasNotice": {
            "@id": "https://example.com/notices/a6f58318-72e6-46a2-bfd7-f36d795e30cd",
            "@type": "dpv:ConsentNotice",
            "dct:date": "2024-01-01",
            "dct:language": "EN",
            "dct:coverage": "2024-01-01/P12M"
        },
        "dpv:hasConsentStatus": [{
            "@type": ["dpv:ConsentGiven", "dpv:ExpressedConsent"],
            "dpv:isIndicatedBy": "dpv:DataSubject",
            "dpv:hasIndicationMethod": "Interaction in App",
            "dpv:isIndicatedAtTime": "2024-01-01"
        }, {
            "@type": "dpv:ConsentWithdrawn",
            "dpv:isIndicatedBy": "dpv:DataSubject",
            "dpv:hasIndicationMethod": "Interaction in App",
            "dpv:isIndicatedAtTime": "2024-04-20"
        }]
      }
    }
\end{minted}

\end{document}